\documentstyle[12pt]{article}
\def\bbox{{\,\lower0.9pt\vbox{\hrule \hbox{\vrule height 0.2 cm
\hskip 0.2 cm
\vrule  height 0.2 cm}\hrule}\,}}

\begin{document}
\setlength{\unitlength}{1mm}
\title{{\hfill {\small } } \vspace*{2cm} \\
A Note on Entanglement Entropy and Conformal Field Theory}
\author{\\
Dmitri V. Fursaev
 \date{ }}
\maketitle
\noindent  {
{\em Joint Institute for
Nuclear Research,
Bogoliubov Laboratory of Theoretical Physics,\\
 141 980 Dubna, Russia}
\\
e-mail: fursaev@thsun1.jinr.ru
}
\bigskip

\begin{abstract}
It is pointed out that
the entanglement entropy of quantum
fields near the horizon of a two-dimensional
black hole can be derived by means of
the conformal field theory. This can be done in a way analogous
to the computation of the entropy of
BTZ black holes.
The important feature of the considered case
is that the degrees of freedom
of the conformal theory are states localized
in the physical space-time.
\end{abstract}

\bigskip

\baselineskip=.6cm


In the last months there was a large interest
to the conjectured correspondence between field theories
on asymptotically anti-de Sitter (AdS) space-times
and conformal field theories (CFT) on the boundary
of AdS \cite{Maldacena}--\cite{Witten}.
In particular,
this correspondence is a key moment for understanding why
the entropy of BTZ black holes \cite{BTZ} can be computed
in the framework of the boundary CFT (see \cite{Martinec},\cite{HKL}
and also references therein).

It should be noted however that in the Strominger's computations
of the entropy \cite{Strominger} CFT is a dual theory.
Consequently, it
does not explain what are the degrees of freedom of the
black hole itself and where are they localized
(for a discussion of this issue see, e.g.,
\cite{Carlip}). To put it in another way, the
equivalence between bulk and boundary theories
is well established
on the level of thermodynamics\footnote{This means that
the Euclidean Liouville
action \cite{Polyakov:87}
induced by CFT on the boundary correctly reproduces the classical free
energy of the BTZ black hole, see, e.g.,
\cite{Martinec}.}, but still
there is no a definite opinion about the microscopic structure of the bulk
gravity corresponding to states of CFT.

The aim of this note is to draw an attention to a situation
when
the degrees of freedom of the conformal field theory
can be microscopic states in the physical space-time.
This example is black holes in two dimensions.
The entanglement entropy of quantum fields which appears
due to the presence of the black hole horizon can
be correctly reproduced in the leading approximation by
means of 2D CFT. This happens for a very simple reason:
near the horizon all quantum fields are
effectively massless and thus conformally invariant.

The fact that CFT can be used for computation of the
entanglement (or geometric) entropy is not new.
This was demonstrated in \cite{HLW} by Holzhey, Larsen, and
Wilczek who suggested several elegant methods
to compute the entropy by
using CFT and the replica method.
We want to show how to come to the same result
in a different way which is parallel to the computations
by Strominger \cite{Strominger} of the BTZ black hole entropy.
As a practical application, our results may be useful
for derivation of the entropy of two-dimensional
black holes (a recent research in this direction can be found
in \cite{CaMi}).

\bigskip\bigskip

To begin with let us recall why the masses
have no effect on the entanglement entropy in the leading
approximation. The exterior region of a static two-dimensional black
hole is described by the metric
\begin{equation}\label{1.1}
ds^2=-g(x)dt^2+g^{-1}(x)dx^2~~~,~~x_h<x\leq x_b~~~,
\end{equation}
where coordinates $x_h$ and $x_b$ correspond to the positions
of the horizon and the boundary, respectively.
On the horizon $g(x_h)=0$.
For non-extremal black holes $g'(x_h)$ is not zero, and one can define the
``surface gravity'' constant $k=g'(x_h)/2$.
Let us consider as an example a scalar field $\phi$
on space-time (\ref{1.1}).
The equation for $\phi$ is
\begin{equation}\label{1.2}
(-\nabla^2+\xi R +m^2)\phi(x)=0~~~.
\end{equation}
It is reduced to a relativistic analog of the Schroedinger equation
for wave functions $\phi_\omega$
of single-particle excitations with frequencies $\omega$.
By making in (\ref{1.2})
the substitution $\phi(t,x)=\exp(-i\omega t)\phi_\omega (x)$
one finds
\begin{equation}\label{1.3}
\bar{H}^2\phi_\omega=\omega^2\phi_\omega~~~,
\end{equation}
\begin{equation}\label{1.4}
\bar{H}^2=-\partial_y^2+g(\xi R+m^2)~~~,
\end{equation}
where coordinates $y$ and $x$ are related as $dy=dx/g$.
As follows from (\ref{1.4}), near the horizon
all mass terms can be
neglected because of the factor $g$ and
the single-particle
Hamiltonian $\bar{H}$ is just $\sqrt{-\partial_y^2}$.
The same property is true for other fields\footnote{A
discussion
of the general situation in four-dimensional space-times can be
found, for instance, in \cite{FF:98a}.}.
In fact, $\bar{H}$ is the Hamiltonian of single-particle excitations
on the ultrastatic space-time
\begin{equation}\label{1.5}
d\bar{s}^2=-dt^2+dy^2~~~,
\end{equation}
which is conformally related to (\ref{1.1}).
In (\ref{1.5}) the position of the horizon is at infinity. Thus,
one is dealing with massless fields on an infinite space. The
entropy of such fields is infrared divergent quantity. By making the size
of the system finite and equal to $ l$
one easily finds the free-energy $F$, energy $E$ and entropy $S$
at the temperature $\beta^{-1}$
\begin{equation}\label{1.6}
F=-{\pi \over 6\beta^2} {l}~~,~~
E={\pi \over 6\beta^2} {l}~~,~~
S={\pi \over 3\beta}{ l}~~~.
\end{equation}
The finite size $ l$ is equivalent to introducing a cutoff near
the horizon at some proper distance $\epsilon$. Usually, to find
the relation
between ${l}$ and $\epsilon$,
 metric (\ref{1.1}) is represented in another form
\begin{equation}\label{1.7}
ds^2=e^{-\varphi}(-\kappa^2\rho^2dt^2+d\rho^2)~~~.
\end{equation}
Then, at small $\epsilon$,
\begin{equation}\label{1.8}
{ l}={1 \over k}\left(-\frac 12 \varphi_h+\ln {\rho_b
\over \epsilon}\right)~~~,
\end{equation}
where $\varphi_h$ is the value of the conformal factor
at the horizon and $\rho_b$ is the boundary value of $\rho$.
The entanglement entropy corresponding to a Hartle-Hawking
vacuum
is evaluated at the Hawking temperature
$\beta_H^{-1}=k/(2\pi)$.
For massless fields, calculated in this way $S$,
see (\ref{1.6}), represents the exact (in the leading
order) result for the
entropy.

\bigskip\bigskip

Let us now show how to carry out the computations
of the entanglement entropy along the lines of
Ref. \cite{Strominger}. We consider $N$ fields, which may be,
for instance, scalars. According to (\ref{1.6}), in the
leading order
\begin{equation}\label{1.9}
E=N {\pi \over 6\beta^2}{ l}~~~,~~~
S=N {\pi \over 3\beta}{ l}~~~.
\end{equation}
The corresponding CFT near the horizon is characterized
by the central charge $c=N$, see \cite{Polyakov:87}.
The relation between the Hamiltonian of the system and
generators of the Virasoro algebra follow from the representation
of the metric (\ref{1.5}) in the form
\begin{equation}\label{1.10}
d\bar{s}^2=\left({{ l} \over \pi}\right)^2(-d\eta^2+dz^2)=
\left({ { l} \over \pi}\right)^2 dudv~~~,
\end{equation}
\begin{equation}\label{1.11}
u={z+\eta \over 2}~~~,~~~v={z-\eta \over 2}~~~.
\end{equation}
Consequently,
\begin{equation}\label{1.12}
\partial_t={\pi \over 2{ l}}(\partial_u-\partial_v)~~~.
\end{equation}
In (\ref{1.10}) the coordinate $z$ ranges from $0$ to $\pi$.
This corresponds to a theory on an interval where the points $z=0$ and $z=\pi$
are independent.
In order to carry out the computations it is convenient
to pass to a theory where $z$ is a periodic coordinate.
This can be done if one considers two equivalent CFT's on the
intervals with the length $\pi$ and makes from
them a CFT on a circle by gluing together the ends of the
intervals.  In the obtained theory
$z$ has the periodicity $2\pi$.

One has two copies of the Virasoro algebra where
the elements $L_n$ and $\bar{L}_n$ can be defined in a standard way,
as the generators of the coordinate transformations,
$\delta u=e^{inu}$ and
$\delta v= e^{inv}$, respectively.
As a result of relation (\ref{1.12}),
the Hamiltonian $H$ of the system which generates transformations
along the Killing time $t$ is represented as
\begin{equation}\label{1.13}
H={\pi \over 2{ l}}(L_0-\bar{L}_0) ~~~.
\end{equation}
Similarly, translations of the
system along $y$ are generated by the momentum
\begin{equation}\label{1.14}
P={\pi \over 2{ l}}(L_0+\bar{L}_0)~~~.
\end{equation}
Because the system is at rest the average momentum
is zero. On the other hand, the average value of $H$
coincides with the energy $E$
in (\ref{1.9}). This fixes the average values $h$ and $\bar{h}$
of $L_0$ and $\bar{L}_0$, respectively. In the given quantum state
\begin{equation}\label{1.15}
h=-\bar{h}={N \over 6}{{ l}^2 \over \beta^2}~~~.
\end{equation}
In the limit when $ l$ is very large ($\epsilon$ goes
to zero), $h \gg 1$ and one can use Cardy's formula
to estimate the degeneracy of $L_0$ and $\bar{L}_0$.
In this approximation the total degeneracy $D$ is
\begin{equation}\label{1.16}
\ln D=2\pi\sqrt{{ch \over 6}}+2\pi\sqrt{{c|\bar{h}| \over 6}}
\end{equation}
and by taking into account that in our case the central charge
$c=N$ we find
\begin{equation}\label{1.17}
\ln D=2N {\pi \over 3}{ l \over \beta}~~~.
\end{equation}
Finally, we have to remember that $D$ is the number of states
of the system with the doubled Hilbert space which results
from the trick with the periodization of the coordinate $z$.
The real number
of states of the
system we are interested in is $\sqrt{D}$.
Thus, the entropy is
\begin{equation}\label{1.18}
S=\frac 12 \ln D
\end{equation}
and it coincides exactly with the required value
in Eq. (\ref{1.9}). To get the entanglement entropy of
the fields in the Hartle-Hawking state
one has to put $\beta=\beta_H$ in (\ref{1.18}).

\bigskip
We have shown how to calculate the entanglement entropy by means
of CFT in the way which is close to computations of the
entropy of three-dimensional BTZ black holes \cite{Strominger}.
The difference between our case and three-dimensional one is
that here the degrees of freedom of CFT are the physical microscopic
states of the theory.
It is worth pointing out that in the above
computations the entropy was determined
by the degeneracy of states at the given energy, i.e., as
the entropy of a microcanonical ensemble. On the other hand,
the results of \cite{HLW} are based on using the replica method
and they are
equivalent to calculation of the entropy of a canonical ensemble.

\section*{Acknowledgments}

I am grateful to G. Cognola, D. Klemm, L. Vanzo, and S. Zerbini for
the discussions and hospitality at Trento.
I also want to thank INFN and the University of Trento
for the financial support of my visit.

\indent

\end{document}